\documentclass[12pt,a4paper]{article}

\newcommand{\ppbar}{p^{\!\!\!\!\!\textsuperscript{\tiny{(--)}}}\!\!}
\newcommand{\ptmiss}{{p\!\!\!\!\! \not \,\,\,\,}_T}
\newcommand{\ptmax}{p_T^{\ell,\mathrm{max}}}
\newcommand{\ptmin}{p_T^{\ell,\mathrm{min}}}
\newcommand{\Esix}{\text{E}_6}
\newcommand{\zp}{{Z'_\lambda}}
\newcommand{\rmu}{r_\mu}
\newcommand{\rtau}{r_\tau}
\newcommand{\ee}{e^\pm e^\pm}
\newcommand{\emu}{e^\pm \mu^\pm}
\newcommand{\mumu}{\mu^\pm \mu^\pm}

\usepackage{graphicx}
\usepackage{amsmath,amssymb}

\usepackage{amsmath,amssymb,graphicx}
\usepackage{epsfig}
\usepackage{cite}

\begin{document}

\begin{center}
\begin{Large}
{\bf Like-sign dilepton signals from a leptophobic \\ $Z'$ boson}

\end{Large}

\vspace{0.5cm}
F. del Aguila, J. A. Aguilar--Saavedra\\[0.2cm] 
{\it Departamento de F\'{\i}sica Te\'orica y del Cosmos and CAFPE, \\
Universidad de Granada, E-18071 Granada, Spain} \\[0.1cm]
\end{center}

\begin{abstract}
A new leptophobic neutral gauge boson $Z'$ with small mixing to the $Z$ can 
have a mass as light as $M_{Z'} \sim 350$ GeV, and still have escaped
detection at LEP and Tevatron. Such a $Z'$ boson can be derived from $\Esix$
and,  if the new heavy neutrino singlets in the $\mathbf{27}$ representation
are lighter than $M_{Z'}/2$, the process
$p\ppbar \to Z' \to NN \to \ell^\pm \ell^\pm X$
is observable. Indeed, this new signal could explain the small excess of
like-sign dileptons found at Tevatron. Implications for LHC are also
discussed. In particular, the Tevatron excess could be confirmed with less
than 1 fb$^{-1}$, and leptophobic $Z'$ masses up to 2.5 TeV can be probed
with 30 fb$^{-1}$.
\end{abstract}

\section{Introduction}

New neutral gauge bosons, generically denoted as $Z'$, arise in a variety of
Standard Model (SM) extensions, including well-known grand-unified models as
$\Esix$ as well as little Higgs and extra dimensional models\cite{Yao:2006px}.
The predicted
$Z'$ bosons typically couple to quarks and charged leptons, thus they can
produce a very clean signal at hadron colliders: a pair of opposite charge
leptons with very high invariant mass and transverse momenta.
Non-observation of this signal at Tevatron has placed limits $M_{Z'} \gtrsim
600-700$ GeV
on $Z'$ bosons appearing in several popular scenarios.
Obviously, if a $Z'$ boson does not couple to charged leptons these constraints
do not apply, and one has to look for the new $Z'$ in other final states.
A striking possibility occurs if decays to heavy Majorana neutrinos
$Z' \to N N$ are kinematically allowed. The subsequent lepton number violating
(LNV) decay $N N \to \ell^\pm W^\mp \ell^\pm W^\mp$, with a
branching ratio around 12.5\% if $N$ is a SM singlet, 
produces two energetic like-sign charged leptons (of different flavour in
general) plus additional jets or
leptons, depending on the $W^\mp W^\mp$ decay mode.
For $M_{Z'}$ up to several hundreds of GeV and $Z' \to NN$
not suppressed by phase space, this process can have a large cross section
already at Tevatron energies, while its backgrounds are relatively small,
especially at Tevatron
\cite{Abulencia:2007rd,D0dimuon}.
Other interesting final state for leptophobic (that is, not coupling to
SM leptons) $Z'$ bosons is $Z' \to
t \bar t$ \cite{Gehrmann:1996qw},
in which an excess might be observed if a $Z'$ boson exists.

In this Letter we focus on like-sign dilepton signals, motivated by an
apparent excess at Tevatron \cite{Abulencia:2007rd}. In the next section we
will show that, unlike many other new physics scenarios,
$Z' \to NN \to \ell^\pm W^\mp \ell^\pm W^\mp$ decays could explain this
like-sign dilepton excess, reproducing the kinematics as well as any
relative number of $\ee$, $\mumu$ and $\emu$ events.
But, even if this excess is not confirmed by additional experimental data,
the study of like-sign dilepton signals, in particular from $Z'$ decays,
remains quite interesting for LHC, as we demonstrate in section 3.
Their backgrounds have moderate
size, arising mainly from: (i) processes with one or two isolated leptons
resulting from $b$ decays, especially $t \bar t nj$ and $b \bar b nj$ production
(where $nj$ stands for $n$ additional jets at parton level);
(ii) processes where extra neutrinos and/or charged leptons are
produced and missed by the detector (mainly $WWnj$, $WZnj$ and $Wt \bar t nj$
production). Thus, like-sign dilepton final states are relatively clean,
and we will
find that they allow to probe leptophobic $Z'$ masses above 2 TeV,
surpassing the sensitivity
of other $Z'$ decay channels such as $Z' \to t \bar t$.
The last section is devoted to summarise our results.

\section{Like-sign dileptons at Tevatron}
\label{sec:2}
The small dilepton excess (44 events for $33.2 \pm 4.7$ expected)
found by CDF \cite{Abulencia:2007rd} might be a statistical fluctuation,
or an uncontrolled systematic error. But, if we put aside these two
hypotheses, it is quite demanding to explain the excess invoking to new
physics. This is because:
\begin{itemize}
\item[(i)] The simplest new physics scenarios giving this signal also lead
to other much larger effects, which have not been found.
\item[(ii)] Even predicting a like-sign dilepton excess, the kinematics must
match the one observed, what is non-trivial. The transverse
momenta distribution of the leading charged
lepton exhibits a rather flat excess
distributed from low to high $p_T$ values, in contrast with most common
processes which sharply concentrate at low $p_T$. Likewise, the $\ell \ell$
invariant mass distribution shows an excess
up to relatively large values $m_{\ell \ell} \sim 160$ GeV.
\end{itemize}
We illustrate these difficulties concentrating on new physics processes
with genuine lepton number
violation.\footnote{New lepton number conserving (LNC) processes, as for example
$W' Z$ and $W Z'$
production with one charged lepton missed by the detector, can give
$\ell^\pm \ell^\pm$  signals as well. However, $W' Z$ and $W Z'$ production
are sub-leading with
respect to $W' (\to \ell \nu)$  and $Z' (\to \ell^+ \ell^-)$ production, which
have not been observed at Tevatron. A supersymmetric interpretation 
of these dilepton events is also disfavoured by the absence of trilepton
signals \cite{Aaltonen:2007sw}.}
The simplest one is the production of a
heavy Majorana neutrino singlet with a charged lepton,
$p\bar p \to W \to \ell N \to \ell^\pm \ell^\pm W^\mp \to \ell^\pm \ell^\pm X$.
(Pair production $p\bar p \to Z \to N N$ is much more suppressed by mixing as
well as by phase space, see for example Ref.\cite{delAguila:2006dx}.)
For the heavy neutrino (singlet)
mixings allowed by present constraints \cite{Bergmann:1998rg,Bekman:2002zk}
the cross section of this process at
Tevatron exceeds a handful of events only for $m_N < M_W$,
when the $W$ is produced on
its mass shell \cite{delAguila:2007em}.
But for $m_N < M_W$ the transverse momenta and invariant mass of the two
leptons are
very small, in contrast with the distributions in Ref.~\cite{Abulencia:2007rd}
which show an excess at larger values.
Another process would be $p \bar p \to W \to NE \to \ell^\pm \ell^\pm X$.
In this case $N$ must transform non-trivially under $\mathrm{SU}(2)_L$
(see for example Ref.~\cite{Bajc:2006ia} for $N,E^\pm$ transforming as a
triplet). This process is mainly suppressed by the $W$ $s$-channel propagator
for large $N,E$ masses.

In order to have a sizeable $N$ production cross section for larger neutrino
masses, say $m_N = 150$ GeV, two obvious possibilities are
to introduce additional charged ($W'$) \cite{Keung:1983uu,Datta:1992qw} or
neutral ($Z'$) gauge bosons.
(A third possibility would be to produce $\ell N$ via the exchange of a
charged scalar, but in the absence of additional interactions the mixing of the
heavy neutrino, mainly a gauge singlet, is very small.)
The first one seems difficult to implement
while keeping agreement with present constraints on new charged interactions.
The new $W'$ must couple to quarks of the first generation in order to have a
sizeable production cross section
at hadron colliders, but not to the electron and muon, so that direct production
limits do not apply.
Besides, the $W'$ boson must be light enough
to be produced, and its coupling cannot be right-handed, otherwise there is a
stringent limit $M_{W'} \gtrsim 2$ TeV from the kaon
mass difference \cite{Barenboim:1996nd}.

In the following we explore the second possibility. Like-sign dileptons from
$Z'$ decays may be produced in any SM extension with an extra $Z'$ boson and
heavy Majorana neutrinos, provided $M_{Z'} > 2 m_N$. In particular, if the
new boson is leptophobic the lower bound on its mass is rather weak, and
sizeable signals are possible already at Tevatron. Early studies on
leptophobic $Z'$ bosons
\cite{Chiappetta:1996km,Altarelli:1996pr,Babu:1996vt,Barger:1996kr}
were motivated by the
initial disagreement between the $Z \to b \bar b$, $Z \to c \bar c$ decay rates
measured at LEP and the SM predictions \cite{LEPrbrc}.
Although the differences disappeared with more LEP data, the possibility of
such a new gauge boson is still interesting by itself. There is a variety of
models with extra leptophobic gauge bosons. 
For simplicity, we will
restrict ourselves to an $\Esix$ model in which heavy Majorana neutrinos
appear naturally \cite{Gursey:1975ki}, but
our conclusions are more general. The neutral interactions
of the standard bosons and the new $Z'$ are
described by the Lagrangian \cite{DelAguila:1995fa}
\begin{equation}
\mathcal{L}_\mathrm{NC} = - \bar \psi \gamma_\mu \left[ T_{3} g W_3^\mu
+ \sqrt{\textstyle \frac{5}{3}} Y g_Y B^\mu + Q' g' Z_\lambda^{'\mu} \right]
 \psi \,,
\end{equation}
where a sum over the three families of $\mathbf{27}$ fermions
in the fundamental $\Esix$ representation is understood.
$Y$ is the SM hypercharge properly normalised, and the
extra charges $Q'$ of the new boson $\zp$
correspond to the only leptophobic combination within
$\Esix$ \cite{delAguila:1986iw,del Aguila:1986ez}
\begin{equation}
Q' = 3/\sqrt{10} (Y_\eta + Y/3) \,,
\label{charge}
\end{equation}
with $Y_\eta$ the extra
$\mathrm{U}(1)$ defined by
flux breaking \cite{Hosotani:1983xw,Hosotani:1983vn,Witten:1984dg}.
For left-handed fields,
\begin{align}
& 2 Q'_u = 2 Q'_{d_1} = Q'_{u^c} = -Q'_{d_2} = -2 Q'_{d_{1,2}^c} =
 - \frac{1}{\sqrt{6}} \,, \nonumber \\
& Q'_{\nu_{1,2}} = Q'_{e_{1,2}} = Q'_{e_1^c} = 0 \,, \nonumber \\
& Q'_{e_2^c} = Q'_{\nu_3} = -Q'_{\nu_{4,5}} = \frac{3}{2 \sqrt{6}} \,  \,.
\end{align}
A detailed discussion of the phenomenological constraints on this SM
extension can be found in Ref.~\cite{Babu:1996vt}, where a nearly-leptophobic
model with $Q' \sim Y_\eta + 0.29 Y$ is studied among several other
alternatives.
This supersymmetric model has the largest field content consistent with
perturbative unification of gauge couplings at the GUT scale. The two points
relevant here and which any such SM extension must satisfy are:
\begin{itemize}
\item[(i)] The $Z-Z'_\lambda$ mixing must be small to maintain the good
agreement with precision electroweak data. This does not pose a problem in
principle, because
this mixing can be made as small as experimentally needed invoking a
cancellation between the contributions of the vacuum expectation values
of the Higgs bosons giving masses to the up and down quarks, respectively.
\item[(ii)] The fermion mass generation mechanism must explain why the extra
fermions are heavier than the SM ones. This can be understood
due to their vector-like character under the SM gauge group. Still, their
masses are protected by the extra gauge interaction and cannot be much
heavier than the extra gauge boson.
\end{itemize}
In these models the neutrino sector is
rather involved and requires a detailed analysis which
will be presented elsewhere.
In each family, one of the two extra neutrino singlets $\nu_{4,5}$ 
can obtain a large mass (a Majorana mass through a
non-renormalisable term, or a heavy Dirac mass if it couples to an
additional $\Esix$ singlet). The three resulting heavy neutrinos $N_i$
(one per family) are the ones we are interested in.
The other three neutrino singlets tend to remain massless, and they can
combine with
the SM neutrinos to form Dirac fermions (in which case the corresponding Yukawa
couplings must be very small or zero). In this phenomenological study
we will assume for simplicity that these latter neutrinos are heavy
enough so that they are not produced in
$Z'_\lambda$ decays. Otherwise, the total $Z'_\lambda$ width
is larger, and in the worst case this would amount to a $\sim 20\%$
decrease of the like-sign dilepton cross sections presented below.

In summary, the extra vector-like lepton doublets and quark singlets of
charge $-1/3$ are assumed to be heavier than $M_\zp/2$, as three of the heavy
neutrinos. Possible supersymmetric partners are taken heavier as well.
On the other hand, the three remaining heavy neutrinos $N$ entering in our
discussion are assumed lighter than $M_\zp/2$.
Their mixing with the light leptons
(see Ref.~\cite{delAguila:2006dx} for details and
notation) can be made of order $V \sim O(10^{-6})$,
small enough to avoid too large contributions to light
neutrino masses. Even for a mixing of this size heavy neutrinos would decay
within the detector.

\begin{figure}[htb]
\begin{center}
\epsfig{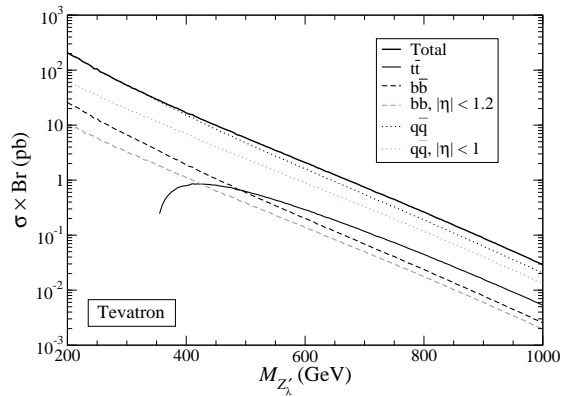}
\caption{Total cross section for $\zp$ production at Tevatron, and cross sections
for several SM final states.}
\label{fig:cross}
\end{center}
\end{figure}

The $\zp$ production cross section at Tevatron (which is obviously independent
of the heavy neutrino masses) is plotted in
Fig.~\ref{fig:cross} as a function of $M_\zp$.
We also plot the maximum ({\em i.e.} when $\zp$ decays
only to SM fermions) cross sections for $t \bar t$, $b \bar b$ and $q \bar q$
final states, also including $q=b$.
The coupling constant of the new
$\mathrm{U}(1)'$ has been fixed for reference as
$g' = \sqrt{5/3}\, g_Y = \sqrt{5/3}\, g \,s_W/c_W$,
and cross sections are calculated using CTEQ5L parton distribution functions
\cite{Lai:1999wy}.
For easier comparison with Tevatron
measurements of the dijet and $b \bar b$ cross sections\footnote{For
the latest results see {\tt http://www-cdf.fnal.gov},
{\tt http://www-d0.fnal.gov}} we also plot the latter two with pseudo-rapidity
cuts. A light $\zp$ might be visible in $b \bar b$ final states, but for $M_\zp
\gtrsim 350$ GeV this seems quite difficult.

\begin{figure}[htb]
\begin{center}
\epsfig{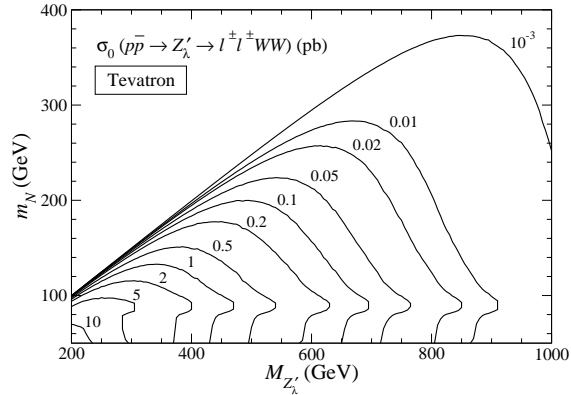}
\caption{Total cross section for $\ell^\pm \ell^\pm W^\mp W^\mp$
production at Tevatron, summing final states with any combination of
$\ell=e,\mu,\tau$.}
\label{fig:con}
\end{center}
\end{figure}

If $\zp \to NN$ decays are kinematically allowed they provide the cleanest
signals of the $\zp$ boson. A heavy Majorana neutrino $N$
can decay to $W^+ \ell^-$,
$W^- \ell^+ $, $Z \nu_\ell$ and $H \nu_\ell$, where $\ell=e,\mu,\tau$,
with partial widths (see for example Ref.~\cite{delAguila:2006dx})
\begin{eqnarray}
\Gamma(N \to W^+ \ell^-) & = &  \Gamma(N \to W^- \ell^+) \nonumber \\
& = & \frac{g^2}{64 \pi} |V_{\ell N}|^2
\frac{m_N^3}{M_W^2} \left( 1- \frac{M_W^2}{m_N^2} \right) 
\left( 1 + \frac{M_W^2}{m_N^2} - 2 \frac{M_W^4}{m_N^4} \right) \,, \nonumber
\\[0.1cm]
\Gamma(N \to Z \nu_\ell) & = &  \frac{g^2}{64 \pi c_W^2} |V_{\ell N}|^2
\frac{m_N^3}{M_Z^2} \left( 1- \frac{M_Z^2}{m_N^2} \right) 
\left( 1 + \frac{M_Z^2}{m_N^2} - 2 \frac{M_Z^4}{m_N^4} \right) \,, \nonumber
\\[0.2cm]
\Gamma(N \to H \nu_\ell) & = &  \frac{g^2}{64 \pi} |V_{\ell N}|^2
\frac{m_N^3}{M_W^2} \left( 1- \frac{M_H^2}{m_N^2} \right)^2 \,.
\label{ec:widths}
\end{eqnarray}
Within any of these four modes, the branching fractions for
individual final states $\ell = e,\mu,\tau$
are in the ratios $|V_{eN}|^2 \,:\, |V_{\mu N}|^2 \,:\, |V_{\tau N}|^2$.
However, as it can be clearly seen from Eqs.~(\ref{ec:widths}),
the total branching ratio for each of the four channels above
(summing over $\ell$)
is independent of the heavy neutrino mixing and determined only by $m_N$
and the Higgs boson mass, fixed here as $M_H = 120$ GeV. 
Then, the total $\ell^\pm \ell^\pm W^\mp W^\mp$ cross section, shown in
Fig.~\ref{fig:con}, only depends
on $M_\zp$ and the three heavy neutrino masses, taken to be equal
for simplicity, $m_{N_i} \equiv m_N$ for $i=1,2,3$. The small bump on the
right part of the lines is caused by the increase in $\text{Br}(N \to W \ell)$
for $M_W \lesssim m_N \lesssim M_Z$.

\begin{figure}[htb]
\begin{center}
\epsfig{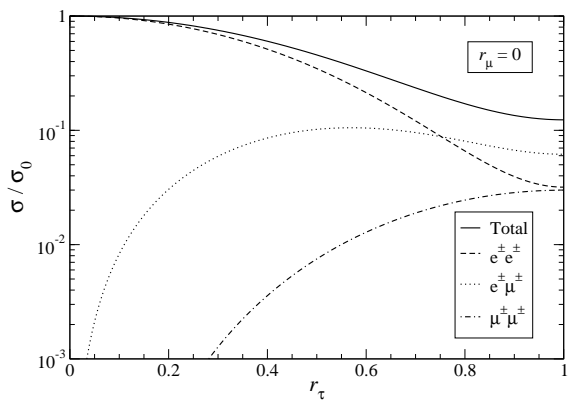}
\caption{Ratio $\sigma/\sigma_0$ between the dilepton cross section to
electron and muon final states and the total one in Fig.~\ref{fig:con}. The
parameters $r_\mu$, $r_\tau$ are defined in Eq.~(\ref{ec:r}).}
\label{fig:r}
\end{center}
\end{figure}

As we have emphasised before,
reproducing the correct kinematics of the apparent like-sign dilepton excess
at CDF is
non-trivial. The presence of events with large missing momentum $\ptmiss$
requires heavy neutrino mixing with the $\tau$, so that decays $N \to \tau W$
with $\tau$ decaying leptonically produce neutrinos in the final state. But the
presence of electrons and/or muons with large transverse momentum
also suggests heavy neutrino mixing with the electron and/or muon. In this
section we do not address
the flavour dependence of the final state (that is, the relative number of
$e^\pm e^\pm$, $\mu^\pm \mu^\pm$ and $e^\pm \mu^\pm$ events) but our main
interest is to reproduce the size and kinematics of the dilepton
excess. The reader can
easily convince
himself that {\em any} relative rate of dielectron, dimuon and
$e^\pm \mu^\pm$ events can be accommodated by choosing adequate mixings
$V_{e N_i}$, $V_{\mu N_i}$ and $V_{\tau N_i}$. Bearing this in mind, one can
reduce the number of free parameters in the analysis.
We assume equal
mixing with the three heavy neutrinos,
$V_{\ell N_i} \equiv V_{\ell N}$,
parameterised as
\begin{eqnarray}
|V_{e N}|    & = & V \cos \frac{\pi}{2} r_\tau \cos \frac{\pi}{2} r_\mu \,,
   \nonumber \\
|V_{\mu N}|  & = & V \cos \frac{\pi}{2} r_\tau \sin \frac{\pi}{2} r_\mu \,,
   \nonumber \\
|V_{\tau N}| & = & V \sin \frac{\pi}{2} r_\tau \,.
\label{ec:r}
\end{eqnarray}
Note that for $V \lesssim 10^{-3}$ constraints from lepton
flavour-violating processes~\cite{Ilakovac:1994kj,Tommasini:1995ii}
and neutrinoless double beta decay~\cite{Aalseth:2004hb,Benes:2005hn}
are automatically
satisfied independently of $\rmu$ and $\rtau$.
For the remaining of this section we take $r_\mu = 0$ (no mixing with the muon)
for simplicity. Then,
the relative mixing with the electron and tau lepton (and thus the branching
ratios for $N \to e W$ and $N \to \tau W$, which are the relevant quantities
for our analysis) depend on a
single parameter $r_\tau$, ranging from 0 to 1.
The values $r_\tau=0$ and $r_\tau=1$
correspond to
$V_{\tau N} = 0$ and $V_{e N} = 0$, respectively, while $r_\tau=0.5$ when both
couplings are equal. The actual dilepton cross section $\sigma$ for final
states with electrons and/or muons can be straightforwardly obtained
in terms of the total cross section $\sigma_0$ in Fig.~\ref{fig:con},
taking into account the branching ratios
for $N \to e W$ and $N \to \tau W$ with subsequent decay
$\tau \to e/\mu bar \nu \nu$.
The rescaling factor $\sigma/\sigma_0$ is shown in Fig.~\ref{fig:r}.
It ranges from unity for $r_\tau=0$ (charged current decays only to electrons)
to $0.12$ for $r_\tau=1$ (only to tau leptons).

We select the values $M_\zp = 500$ GeV, $m_N = 150$ GeV
to illustrate
how the new $\ell^\pm \ell^\pm X$
signal can account for the small CDF
excess. These values are chosen so that decays $\zp \to NN$ and $N \to W \ell$
are not too close to threshold. With these parameters we have
$\mathrm{Br}(\zp \to NN) = 0.20$ (summing over the three neutrinos),
$\mathrm{Br}(N \to W^+ \ell^-) = \mathrm{Br} (N \to
W^- \ell^+) = 0.33$, $\mathrm{Br}(N \to Z \nu_\ell) = 0.29$, $\mathrm{Br} (N \to
H \nu_\ell) = 0.05$.
We generate
events using the exact matrix element for
$p \ppbar \to \zp \to NN \to \ell^\pm W^\mp \ell^\pm W^\mp \to \ell^\pm f
\bar f \ell^\pm f \bar f$, including all finite width
and spin effects. Tau leptonic decays are simulated
using the tree-level matrix element. In this section we restrict ourselves to
hadronic decays of the $W$ pair, which amount to 44\% of the total $WW$ decay
branching ratio. We require the same kinematical
cuts on leptons as in Ref.~\cite{Abulencia:2007rd}: (i) transverse momenta
$\ptmax \geq 20$ GeV, 
$\ptmin \geq 10$ GeV, where $\ptmax$ and $\ptmin$ refer to the leading and
sub-leading lepton, respectively; (ii) pseudorapidity $|\eta^{\ell}| \leq 1$;
(iii) dilepton invariant mass larger than 25 GeV.
Additionally, for charged lepton isolation we require a minimum lego-plot
separation $\Delta R \geq 0.4$ between them and also
between them and final state quarks.
\begin{figure}[t]
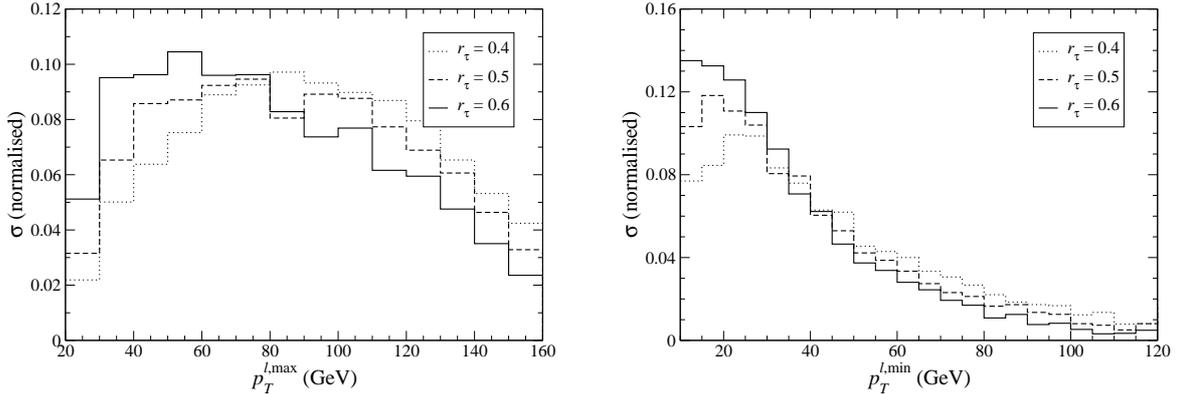

\begin{center}
\begin{tabular}{ccc}
\epsfig{file=Figs/ptmax.eps,height=5.2cm,clip=} & &
\epsfig{file=Figs/ptmin.eps,height=5.2cm,clip=} 
\end{tabular}
\caption{Transverse momenta of the leading ($p_T^{\ell,\mathrm{max}}$)
and sub-leading ($p_T^{\ell,\mathrm{min}}$) charged leptons.}
\label{fig:dist}
\end{center}
\end{figure}
\begin{figure}[t]
\begin{center}
\epsfig{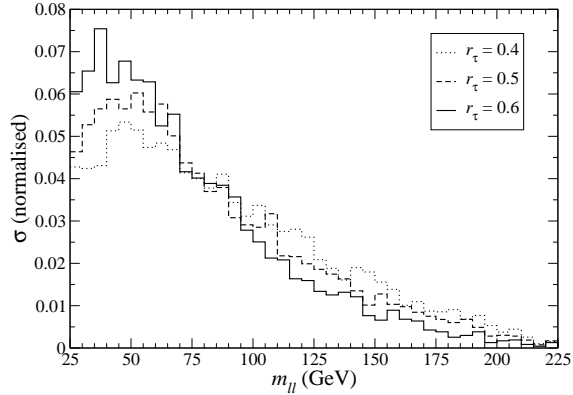}
\caption{Dilepton invariant mass.}
\label{fig:dist2}
\end{center}
\end{figure}
The parton-level distributions for the
transverse momenta of the two leptons are shown in Fig.~\ref{fig:dist},
for three representative values of the mixing parameter $r_\tau$.
It is especially remarkable the slow decrease of the
$\ptmax$ distribution, which is an unusual behaviour (compared for example
with $\ptmin$)
and is in good agreement with the $\ptmax$ distribution of the dilepton excess
in Ref.~\cite{Abulencia:2007rd}.
The dilepton invariant mass is presented in Fig.~\ref{fig:dist2}.
Detector effects are not expected to change
drastically the parton-level predictions for charged lepton momenta but,
unfortunately, the missing energy of the event cannot be reliably estimated
with a parton level analysis. In any case,
we find a slowly decreasing $\ptmiss$
distribution up to $\sim 120$ GeV.
The number of events expected for
$r_\tau=0.4$, $0.5$ and $0.6$ is 21, 14 and 9, respectively, of the same order
of the CDF excess (11 events). An additional signal contribution is expected
from the semileptonic (44\% of the total branching ratio) and dileptonic (11\%)
decays of the
$WW$ pair, when the extra leptons are missed by the detector or have small
energy. Trilepton signals are also present, but a factor $\sim 5$
smaller after the selection criteria used in typical analyses \cite{CDFtri}.
For larger values of $g'$, as obtained from renormalisation group
evolution~\cite{Babu:1996vt}, cross sections scale accordingly.

\begin{figure}[htb]
\begin{center}
\epsfig{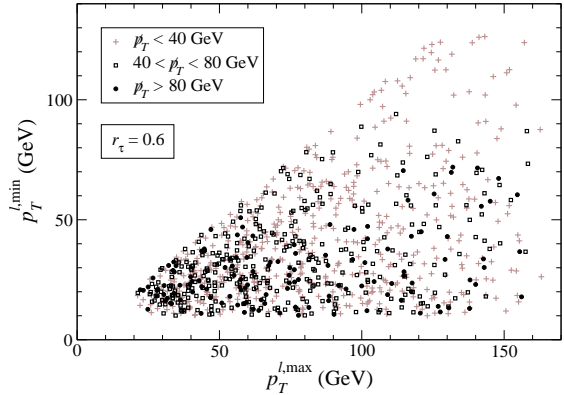}
\caption{Transverse momenta of the leading and sub-leading leptons of an
arbitrary sample of 1000 unweighted $\ell^\pm \ell^\pm q \bar q q \bar q$
events.}
\label{fig:pt12}
\end{center}
\end{figure}

We can also extract interesting information about the signal by representing,
for each event within a typical sample, the
values of $\ptmax$ and $\ptmin$ in a
two-dimensional diagram. This is done in Fig.~\ref{fig:pt12}, using 1000
unweighted $\ell^\pm \ell^\pm q \bar q q \bar q$
events at the partonic level and taking $r_\tau=0.6$, for which
the distributions in Fig.~\ref{fig:dist} resemble most 
the kinematics of the CDF excess. Events are classified according to their
parton-level $\ptmiss$, which illustrates to some extent the missing energy
expected in a real detector. The point density may be interpreted in
terms of probability. We notice that the individual events described in
Ref.~\cite{Abulencia:2007rd} fit well in this distribution. These are:
\begin{enumerate}
\item[(1)] Two electrons with $p_T = 107$ GeV and $p_T = 103$ GeV,
$\ptmiss = 25$ GeV and an additional non-isolated positron with
$p_T = 5$ GeV. This is the event with largest transverse energy.
\item[(2)] Two positrons with $p_T = 73$ GeV and $p_T = 41$ GeV,
$\ptmiss = 96$
GeV. This is the event with second largest transverse energy.
\item[(3)] A $\mu^+$ with $p_T = 66$ GeV, an $e^+$ with $p_T = 10$ GeV
and $\ptmiss = 37$ GeV.
\end{enumerate}

Finally, we note that like-sign dileptons are not the only signature of a
leptophobic $\zp$ boson at Tevatron. Another interesing final
state is given by $\zp \to t \bar t$ decays, which would show up as a 
bump in the $t \bar t$ invariant mass spectrum. 
Within the scenario
described above, the cross section for $p\bar p \to \zp \to t \bar t$ at
Tevatron
is 537 fb. Assuming the same acceptance times efficiency (1.5\%) as in the
latest CDF search for $t \bar t$ resonances \cite{CDFtt},
this would correspond to 8 additional $t \bar t$ events with $m_{t \bar t}$
around 500 GeV for a luminosity of 1 fb$^{-1}$. It is amusing to observe
that a small excess of $t \bar t$ events
has been found  within a sample of 955 pb$^{-1}$ around this region
\cite{CDFtt}, although its statistical significance is of only $\sim 1\sigma$.
The latest D0 analysis \cite{D0tt} with 370 pb$^{-1}$ does not observe any
excess.

\section{Like-sign dileptons at LHC}
\label{sec:3}

\begin{figure}[htb]
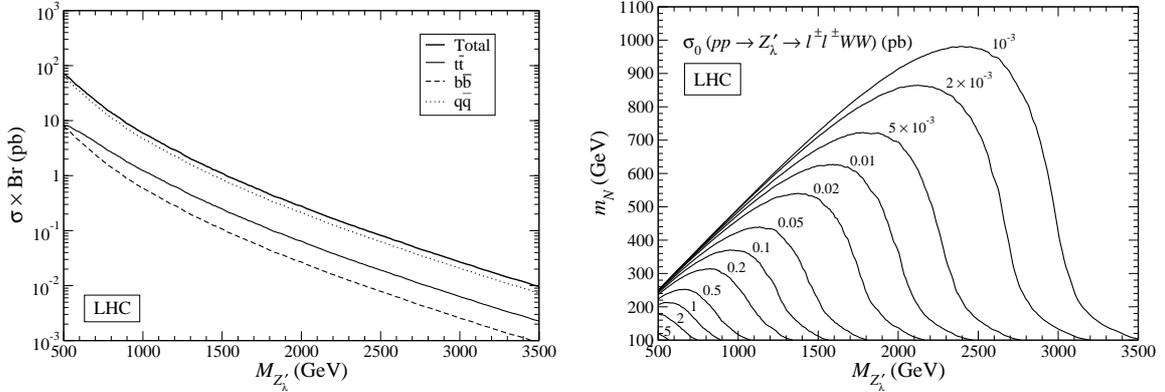

\begin{center}
\begin{tabular}{cc}
\epsfig{file=Figs/cross-lhc.eps,height=5.2cm,clip=} &
\epsfig{file=Figs/contour-lhc.eps,height=5.2cm,clip=}
\end{tabular}
\caption{Left: Total cross section for $\zp$ production at LHC, and cross sections
for several SM final states. Right: Total cross section for
$\ell^\pm \ell^\pm W^\mp W^\mp$ production at LHC, summing final states
with any combination of $\ell=e,\mu,\tau$. }
\label{fig:cross2}
\end{center}
\end{figure}

Independently of whether the CDF dilepton excess is confirmed or not,
like-sign dilepton final states offer an interesting possibility for the
study of leptophobic $Z'$ bosons at LHC.
Discovery of these particles in hadronic
final states is quite difficult, and restricted to relatively low masses
for which the cross sections, plotted in Fig.~\ref{fig:cross2}
(left), are very large. As in Fig.~\ref{fig:cross}, in this plot we
have assumed that $\zp$ decays only to
SM particles, so that the plotted cross sections for SM final states are the
largest ones possible. Let us consider for example the 
$t \bar t$ decay channel. Simulations performed
in Ref.~\cite{pallin} have obtained, assuming a generic resonance $Y$ with
arbitrary mass $m_Y$, the minimum cross section
$\sigma(pp \to Y \to t \bar t)$ for
which $Y$ can be observed with $5 \sigma$. Comparing the data
in Fig.~\ref{fig:cross2} with the results in that  analysis,
it is found that $\zp$
masses up to $\sim 700$ GeV could be discovered in $t \bar t$ final states
with a luminosity of 30 fb$^{-1}$.
Dijet final states, which have the largest $\zp$ decay branching ratio
(see Fig.~\ref{fig:cross2}), have huge backgrounds
and a $Z'_\lambda$ boson would not
be visible in this channel \cite{gumus}. On the other hand,
the cross section for like-sign dilepton production
$pp \to Z'_\lambda \to NN \to \ell^\pm \ell^\pm W^\mp W^\mp$,
presented in Fig.~\ref{fig:cross2} (right) is large in wide areas of the
parameter space, and its backgrounds are much smaller. As it will be shown
below, this process can provide positive signals of a
$Z'_\lambda$ in regions where the hadronic channels cannot achieve enough
statistical significance.

Let us first consider, for the sake of comparison, the scenario with $M_\zp
= 500$ GeV, $m_N = 150$ GeV from the previous section. The larger centre of mass
energy and luminosity at LHC will allow to quickly confirm or discard
the hypothesis of a $\zp$ boson and heavy neutrinos with these masses.
We have performed a fast simulation
of this signal for various values of the heavy neutrino mixings,
parameterised by $\rmu$ and $\rtau$. All decay channels of the $WW$ pair are
included. The SM dilepton backgrounds are taken
from Ref.~\cite{delAguila:2007em} (see this reference for further details).
We require as pre-selection:
\begin{itemize}
\item[(i)] two like-sign isolated charged leptons with 
pseudorapidity $|\eta^\ell| \leq 2.5$ and transverse momentum $p_T^\ell$
larger than 10 GeV (muons) or 15 GeV (electrons), and no additional isolated
charged leptons;
\item[(ii)] no additional non-isolated muons; 
\item[(iii)] at least two jets with $|\eta^j| \leq 2.5$ and $p_T^j \geq 20$
GeV, and no $b$-tagged jets. Although the signal has four jets at the partonic
level, for large $M_\zp$ and $m_N$ it is convenient to require only two jets,
in order to keep the signal as large as possible.
\end{itemize}
With these pre-selection criteria the semileptonic and dileptonic
$WW$ decay channels contribute an extra $\sim 20\%$ to the signal, when the
additional charged lepton(s) are missed by the detector.

The signal has different kinematics depending on $\rtau$: for $\rtau = 0$ the
charged leptons are much more energetic,
while for nonzero $\rtau$ and specially for $\rtau = 1$
the final state has neutrinos from tau decays 
and large missing energy. Here we do not try to optimise the signal
significance in the different channels, but instead we reduce
backgrounds using very simple cuts on lepton transverse momenta,
\begin{equation}
p_T^{\ell,\text{max}} > 30~\text{GeV}\,, \quad
p_T^{\ell,\text{min}} > 20~\text{GeV} \,.
\label{ec:cuts1}
\end{equation}
The number of signal and background events for 1 fb$^{-1}$ is collected
in Table~\ref{tab:nsb1}. Smaller backgrounds are not shown separately but
they are included in the figures in the last row.
In most cases the signal significance (assuming a
20\% background uncertainty) is much larger than $5\sigma$. If the heavy
neutrinos only couple to tau leptons the luminosity required for the discovery
is larger (and cut optimisation is needed).

\begin{table}[htb]
\begin{center}
\begin{tabular}{lcccclccc}
                         & $\ee$ & $\mumu$ & $\emu$
          & \quad \quad  &  & $\ee$ & $\mumu$ & $\emu$ \\
$\zp$ (0,0)     & 923.8 & $-$     & $-$ & &
$t \bar t nj$   & 44.4  & 1.2     & 40.7 \\
$\zp$ (1,0)     & $-$   & 664.1   & $-$ & &
$b \bar b nj$   & 27    & 1       & 9 \\
$\zp$ (0,1)     & 6.1   & 4.4     & 10.3 & &
$tj$            & 1.0   & 0.0     & 1.1 \\
$\zp$ (0.5,0)   & 230.0 & 166.9   & 388.0 & &
$W b \bar b nj$ & 1.6   & 0.0     & 1.4 \\
$\zp$ (0,0.5)   & 161.4 & 4.4     & 52.0 & &
$W t \bar t nj$ & 0.7   & 0.4     & 1.1 \\
$\zp$ (1,0.5)   & 5.9   & 117.5   & 50.5 & &
$WWnj$          & 2.3   & 2.0     & 4.3 \\
& & & & & $WZnj$                   & 6.3   & 3.5     & 9.5 \\
& & & & & $WWWnj$                  & 0.8   & 0.8     & 1.6 \\
& & & & & Total Bkg.               & 84.8  & 9.1     & 69.6 \\
\end{tabular}
\caption{Number of $\ell^\pm \ell^\pm jj$ events at LHC for 1 fb$^{-1}$,
after the cuts in Eqs.~(\ref{ec:cuts1}).
The signal is evaluated assuming $M_\zp = 500$ GeV, $m_N = 150$ GeV,
with the parameters between parentheses standing for $\rmu$ and $\rtau$,
respectively.}
\label{tab:nsb1}
\end{center}
\end{table}

Several remarks regarding these results are in order.
Backgrounds with charged leptons from $b$
decays are large, especially in $\ee jj$ and $\emu jj$ final states
\cite{delAguila:2007em}, and with the loose cuts in Eqs.~(\ref{ec:cuts1})
they remain dominant in these two channels. The number of $\mumu jj$ events
from $WZnj$ production is smaller than the number of $\ee jj$ events due to
the requirement of no non-isolated muons. Therefore, the highest sensitivity
is achieved if heavy neutrinos only couple to the muon, so that decays
$NN \to \mumu W^\mp W^\mp$ have the largest branching ratio possible,
around 12.5\%.

\begin{figure}[htb]
\begin{center}
\epsfig{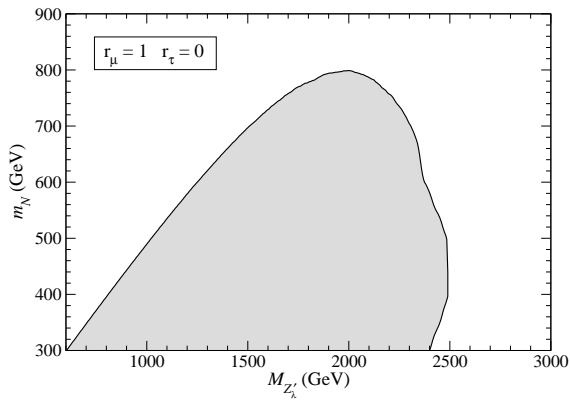}
\caption{$5 \sigma$ discovery limit for $pp \to \zp \to NN$ giving
$\mumu X$ final
states at LHC, for a luminosity of 30 fb$^{-1}$. Heavy neutrinos
are assumed to couple only to the muon.}
\label{fig:5sig}
\end{center}
\end{figure}

For larger $\zp$ masses the charged leptons produced in its decay are more
energetic. This fact can be exploited by requiring large transverse momenta
({\em e.g.} 200 GeV for the leading and 50 GeV for the sub-leading lepton)
and dilepton invariant mass
to reduce backgrounds significantly. In particular, processes
in which one or two charged leptons come from $b$ decays
($t \bar t nj$, $b \bar b nj$, etc.) can be practically removed so that
the numbers of $\ee jj$ and $\mumu jj$ background events are practically
equal. Therefore, for larger $\zp$ masses the sensitivities
in the $\ee jj$ and $\mumu jj$ channels become very similar.
In Fig.~\ref{fig:5sig}
we plot the $5\sigma$ discovery limits in the case that the heavy neutrinos
only couple to the muon, $r_\mu = 1$, $r_\tau = 0$, for a luminosity of
30 fb$^{-1}$. 
The shaded area corresponds to masses $(M_\zp,m_N)$ for which the statistical
significance is larger than $5\sigma$.
It has been obtained generating samples for several points $(M_\zp,m_N)$
close to the boundary.
Performing the corresponding analyses the number of signal events after
selection cuts can be obtained for each point, and interpolation or
extrapolation is used for the remaining points
in the boundary. Notice that this boundary
has not the same shape as the lines of constant cross section in
Fig.~\ref{fig:cross2} (right), because the efficiency after cuts
varies with the $\zp$ and heavy neutrino masses. For $M_\zp \gg m_N$ the
efficiency significantly decreases because the $N$ decay products are very
collinear.

For heavy neutrinos coupling only to the electron the discovery limits are
very similar to those in Fig.~\ref{fig:5sig}. For large $M_\zp$ (upper-right
part
of the boundary) this is because $\ee jj$ and $\mumu jj$ backgrounds have
similar size after cuts. For smaller $M_\zp$ (left part of the boundary)
the discovery limit is determined by the kinematical limit for $\zp \to NN$,
and, even though $\ee jj$ backgrounds are larger in this region,
the signal cross section varies rapidly with $m_N$ and the lines 
for both final states lie very close.

\begin{figure}[htb]
\begin{center}
\epsfig{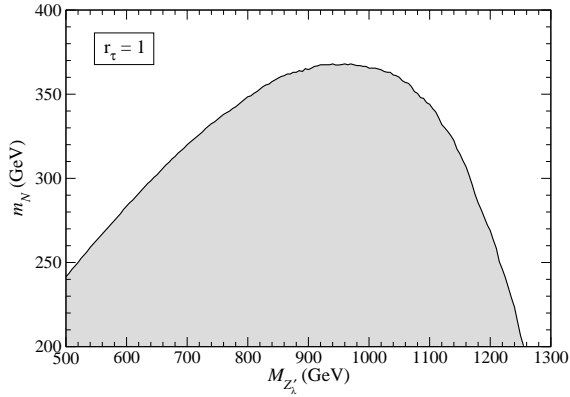}
\caption{$90\%$ exclusion region (shaded area) on $M_\zp$ and $m_N$
if like-sign dilepton signals are not observed at LHC with
a luminosity of 30 fb$^{-1}$.
Heavy neutrinos are conservatively assumed to couple only to the tau.}
\label{fig:90lim}
\end{center}
\end{figure}

If a positive signal is not found at LHC, limits on $M_\zp$ and $m_N$ can
be set. The most conservative limits are obtained assuming that heavy
neutrinos only couple to the tau lepton. If no excess is observed in the
like-sign dilepton channels, the shaded region shown in
Fig.~\ref{fig:90lim} can be excluded at 90\% CL.
This region is obtained by simulating several signal samples and optimising the
kinematical cuts in each case. The relevant variables are the jet multiplicity
(for $(M_\zp,m_N)$ masses not very large it is convenient to require four jets
in event selection), the charged lepton
momenta $\ptmax$ and $\ptmin$, their invariant mass $m_{\ell \ell}$, the
rapidity and azimuthal angle differences, $\Delta \eta_{\ell \ell}$ and
$\Delta \phi_{\ell \ell}$
respectively, the momentum of the most energetic jet $p_T^\text{max}$ and the
missing energy $\ptmiss$. 

\section{Summary}

Like-sign dileptons are interesting final states in which to look for
new physics at hadron colliders. Their backgrounds, mainly from $t \bar t nj$
and $b \bar b nj$ production, have moderate size in contrast with other
LNC final states \cite{delAguila:2007em}. Like-sign
dilepton signals are characteristic of Majorana fermions (such as
new heavy neutrinos) and of doubly charged scalars, both mediating LNV
interactions. They can also appear from LNC processes when additional leptons
are missed by the detector.

Motivated by an apparent like-sign dilepton excess at Tevatron,
we have studied a model in which heavy neutrino pairs can be produced
at hadron colliders via the exchange of an $s$-channel leptophobic
$\zp$ boson. Constraints on the latter are rather loose, and if $Z-\zp$
mixing is negligible the new boson could be as light as $M_\zp \sim 350$
GeV, with very large production cross sections at hadron colliders and,
in particular, leading to potentially large like-sign dilepton signals.
In case that the Tevatron excess is confirmed with more statistics,
a possible explanation might be the one proposed here: a new
$\zp$ boson decaying to heavy neutrino pairs,
$p \bar p \to \zp \to NN \to \ell^\pm \ell^\pm W^\mp W^\mp$.
We have shown that not only the size of the excess but also its kinematics
can be explained with an addtional $\zp$ boson. Taking,
for example, masses $M_\zp = 500$ GeV, $m_N = 150$ GeV,
the $p_T$ distribution of the leading and sub-leading charged leptons and
the dilepton invariant mass can be well accommodated.

As we have already noted, like-sign dilepton signals at hadron colliders
are also predicted in several other SM extensions involving heavy neutrinos.
It is worth comparing the $\zp$ model studied here with some of them.
The most popular ones are: 
\begin{itemize}
\item[(i)] Models with heavy neutrino singlets (as those appearing in type-I
seesaw\footnote{For a recent review on seesaw models of neutrino masses and
their low energy effects see Ref.~\cite{Abada:2007ux}.})
without extra interactions, that is, without $W'$ or $Z'$ bosons.
In this case the main production process is
$p \ppbar \to W \to \ell N$ and
dilepton cross sections are much smaller because they are
proportional to the square of the heavy neutrino mixing with the SM fermions,
which is experimentally constrained to be very small
\cite{Bergmann:1998rg,Bekman:2002zk}.
If $m_N > M_W$, the cross section is also suppressed by the off-shell
$W$ propagator.

\item[(ii)] Models with heavy neutrinos in $\mathrm{SU}(2)_L$ lepton triplets,
as those appearing in type-III seesaw. In this case
heavy lepton pairs can be produced through $s$-channel $W$ boson exchange,
$p \ppbar \to W \to NE$, giving the same final
state studied in this work \cite{Bajc:2006ia}.
The $WNE$ coupling has gauge strength with
mixing $O(1)$ but the cross section is still suppressed by the off-shell $W$
propagator.

\item[(iii)] Left-right models or, more generally, models with an extra
$W'$ and right-handed neutrinos. The latter can be produced in association with
a charged lepton, $p \ppbar \to W' \to \ell N$ \cite{Keung:1983uu,Datta:1992qw},
with interactions of gauge strength.
The cross section is only suppressed by present lower bounds on the $W'$ mass
resulting from direct searches and indirect limits.
\end{itemize}
In these three cases the allowed like-sign dilepton cross sections are
generically smaller than for a leptophobic $\zp$ boson. Hence,
producing large signals at Tevatron, so as to explain the apparent CDF excess,
seems difficult in these models.

Finally, even if the Tevatron excess is diluted with additional
data, like-sign dilepton signals, either from a leptophobic $\zp$ boson
or within the models (i-iii) listed above, 
will remain quite interesting for 
LHC~\cite{delAguila:2007em,Ferrari:2000sp,Gninenko:2006br}.
In the SM extension studied in
this
work, leptophobic $\zp$ bosons will be probed up to masses
$M_\zp \simeq 2.5$
TeV and $m_N \simeq 800$ GeV for a luminosity of 30 fb$^{-1}$,
in the most favourable case that heavy neutrinos do not couple to the tau
lepton.
This $M_\zp$ scale is much higher than the one which
can be probed in the hadronic final states, approximately 700 GeV in
$\zp \to t \bar t$. On the other hand, if a dilepton excess is not
found at LHC, useful limits could be set on the mass of a new $\zp$ boson,
which is, as we have emphasised before, loosely constrained at present.

\vspace{1cm}
\noindent
{\Large \bf Acknowledgements}
\vspace{0.3cm}

\noindent
We thank N. Castro and M. P\'erez-Victoria for useful comments.
This work has been supported by MEC project FPA2006-05294 and
Junta de Andaluc{\'\i}a projects FQM 101 and FQM 437.
J.A.A.-S. acknowledges support by a MEC Ram\'on y Cajal contract.

\end{document}